\documentclass{cambridge6A}
\usepackage{amsfonts,slashed}
\usepackage{arydshln}
\usepackage[greek,english]{babel}
\usepackage{cancel}
\usepackage{upgreek}
\usepackage{float}
\usepackage{url}
\usepackage{latexsym}
\usepackage{amsfonts}
\usepackage{amsmath}
\usepackage{epsfig}
\usepackage{latexsym,amssymb}
\usepackage{amsmath,amssymb,amsthm}
\usepackage{hyperref}
\usepackage{fontenc}
\usepackage{graphicx,tikz}
\usepackage{multirow}
\usepackage{xfrac}
\usepackage{textcomp}
\usepackage{lmodern}
\usepackage{stmaryrd}
\usepackage{cite}

\usepackage{graphicx,tikz}

\renewcommand\thesection{\arabic{section}}
\numberwithin{equation}{section}
\renewcommand{\thefigure}{\arabic{figure}}

\begin{document}

\chapter*{{\bf Exceptional field theories}}

\vspace*{-3.6cm}

\begin{center}
    {\Large Henning Samtleben} \\[2ex]
{\em\small     ENS de Lyon, CNRS, LPENSL, UMR5672, 69342, Lyon cedex 07, France\\[1ex]
Institut Universitaire de France (IUF)\\[2.5ex]
}

{\em\small Invited contribution to ``Half a century of Supergravity''\\
eds. A. Ceresole and G. Dall'Agata (Cambridge Univ. Press, to appear)}

\end{center}

\begin{abstract}
We review exceptional field theories as the duality-covariant reformulation of maximal supergravity theories in ten and eleven dimensions, that make the underlying exceptional symmetries explicit. Beyond their structural role in unifying the various maximal supergravities, we illustrate how they also provide access to very efficient techniques for tackling concrete computational problems in supergravity.
\end{abstract}

\bigskip

\section{Introduction}

Among supergravity's most striking features is the appearance of exceptional symmetry groups upon toroidal reduction of the higher-dimensional theories \cite{Cremmer:1979up}. Maximal supergravities, obtained from compactification of eleven-dimensional and IIB supergravity to lower dimensions exhibit a surprising enhancement of their global symmetry group beyond the geometric symmetries  descending from higher-dimensional diffeomorphisms and tensor gauge transformations. Generalizing the mechanism of the Ehlers symmetry, discovered earlier in general relativity  \cite{Ehlers:1957}, these hidden symmetries enhance the global symmetry group of maximal supergravity in $D$ dimensions to the exceptional group ${\rm E}_{d(d)}$, for $d=11-D$. In particular, the scalar fields in these theories parametrize coset spaces ${\rm E}_{d(d)}/{\rm K}({\rm E}_{d(d)})$, where ${\rm K}({\rm E}_{d(d)})$ is the maximal compact subgroup of the non-compact split form ${\rm E}_{d(d)}$ of the exceptional group $E_d$. These exceptional symmetries were later understood as the supergravity manifestations of the (discrete) U-duality symmetries of string and M-theory \cite{Hull:1994ys}. From the perspective of conventional Riemannian geometry however, these symmetries appear rather mysterious.

Gauged maximal supergravities were first introduced in the 1980s, combining maximal supersymmetry with non-abelian gauge symmetry upon promoting part of the global exceptional symmetries to local gauge symmetries \cite{deWit:1982bul}. 
In particular, in $D=4$ dimensions, the maximal ${\rm SO}(8)$-gauged supergravity describes the consistent truncation of $D=11$ supergravity on a round seven-sphere \cite{deWit:1984nz}.
Later, such deformations were systematically classified and constructed by virtue of
the embedding tensor formalism \cite{Nicolai:2000sc,deWit:2002vt,deWit:2004nw}. 
In this formalism, the embedding tensor which triggers the deformation is treated as a spurionic object transforming in a particular representation of the exceptional symmetry group.
The consistency conditions of the deformation such as supersymmetry and the closure of the gauge algebra can be encoded by algebraic constraints on this tensor.
In case the gauged supergravity admits a higher-dimensional uplift, the components of the embedding tensor can be identified with the parameters of the internal manifold, the background fluxes, the torsion, etc.. Even though every gauging breaks the exceptional symmetry group of the ungauged theory,
the exceptional symmetry structures still play a crucial role in determining the viable deformations,
and the specific couplings of the gauged theory. 

Exceptional field theories (ExFTs) provide a modern unified framework which naturally embeds both, ungauged and gauged supergravities, together with their higher-dimensional ancestors. They amount to a reformulation of eleven-dimensional and IIB supergravity which renders manifest the exceptional structures in higher dimensions before dimensional reduction \cite{Hohm:2013pua}. Upon splitting the higher-dimensional spacetime into an internal and an external $D$-dimensional space, the theory is cast into a form that ressembles the general form of a maximal gauged supergravity in $D$ dimensions. All fields however still depend on all ten or eleven coordinates, and the role of the embedding tensor's structure is taken on by the algebra of generalized diffeomorphisms on the internal space. The latter symmetries combine the internal diffeomorphisms and tensor gauge transformations, realized on an extended geometry \cite{Hull:2007zu,Pacheco:2008ps,Hillmann:2009pp,Berman:2010is,Coimbra:2011ky,Berman:2012vc,Coimbra:2012af,Aldazabal:2013mya,Cederwall:2013naa}.
Notably, the exceptional symmetries are realized as structure groups of the generalized diffeomorphisms.
Interestingly, in this duality covariant framework, the bosonic sector of maximal supergravity 
is uniquely determined by imposing invariance under generalized diffeomorphisms.
Although its fermionic sector can be accommodated within the framework, the exceptional symmetries have, in a sense, 
taken over the role of supersymmetry in determining all the couplings of the bosonic theory.
Beyond their conceptual significance, exceptional field theories have also served as powerful tools in the construction of consistent truncations \cite{Lee:2014mla,Hohm:2014qga} and the computation of Kaluza-Klein spectra \cite{Malek:2019eaz} of maximal supergravity.

\begin{figure}[tb]
\center
\includegraphics[scale=.52]{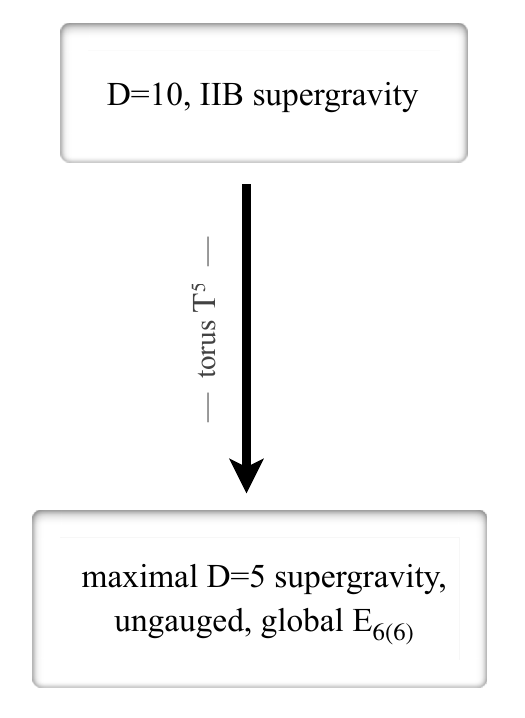}
\caption{
Exceptional symmetry in maximal supergravity}
\label{fig:ungauged}
\end{figure}

A detailed construction of the exceptional field theories for the various exceptional groups has been given in \cite{Hohm:2013vpa,Hohm:2013uia,Hohm:2014fxa,Hohm:2015xna,Abzalov:2015ega,Musaev:2015ces,Berman:2015rcc,Bossard:2018utw,Bossard:2021jix,Bossard:2021ebg}, 
reviews include \cite{Baguet:2015xha,Berman:2020tqn,Sterckx:2024vju}. 
In this chapter, we will rather illustrate these structures at work 
for one characteristic example, the exceptional group E$_{6(6)}$,
and review its role in maximal supergravity within the framework of exceptional field theory.
Specifically, in sections~\ref{sec:ungauged} and \ref{sec:gauged}, we review the structure of ungauged and gauged maximal supergravity in $D=5$ dimensions.
In section~\ref{sec:exft}, we describe the E$_{6(6)}$ exceptional field theory  and its role in bringing together the preceding constructions and their higher-dimensional uplifts. Schematically, the interconnections are depicted in Figures~\ref{fig:ungauged}, \ref{fig:gauged}, and \ref{fig:exft}.
Finally, in section~\ref{sec:KK}, we briefly review an important application of exceptional field theories, serving as a very efficient tool for the computation of Kaluza-Klein spectra around AdS backgrounds with and without supersymmetry.

\section{Maximal supergravity: exceptional symmetries}
\label{sec:ungauged}

Cremmer and Julia in their seminal work \cite{Cremmer:1979up} revealed the emergence of the exceptional symmetry group ${\rm E}_{7(7)}$ in the toroidal reduction of $D=11$ supergravity. This set the stage for the construction of all maximal supergravities in lower dimensions upon combining the respective exceptional global symmetries with maximal supersymmetry. Here, we briefly review the structure of maximal $D=5$ supergravity, based on the exceptional group ${\rm E}_{6(6)}$, first constructed in  \cite{Cremmer:1980gs}. The bosonic field content of the theory comprises the metric ${\rm g}_{\mu\nu}$, $\mu=0, \dots,4$, together with 27 vector fields $A_\mu{}^M$, $M=1, \dots, 27$, and 42 scalar fields. The latter are conveniently described by a symmetric ${\rm E}_{6(6)}$-valued matrix ${M}_{MN}$, parametrizing the coset space ${\rm E}_{6(6)}/{\rm USp}(8)$. The dynamics of the bosonic sector is described by the Lagrangian
\begin{align}
 {\cal L}_{\rm sugra} =\,& \sqrt{-{\rm g}}\,\Big(R
 +\tfrac{1}{24}\,\partial_{\mu}{M}^{MN}\,\partial^{\mu}{M}_{MN}
-\tfrac{1}{4}{M}_{MN}{F}^{\mu\nu M}{F}_{\mu\nu}{}^N\Big)
\nonumber\\
&{}
+{\cal L}_{\rm top} \,,
\label{eq:D5}
\end{align}
where ${M}^{MN}$ denotes the matrix inverse to ${M}_{MN}$, and ${F}_{\mu\nu}{}^M$ is the abelian field strength of $A_\mu{}^M$\,. The topological term of the Lagrangian is a $D=5$ Chern-Simons term defined by
\begin{equation}
d{\cal L}_{\rm top}
= \tfrac{1}{6}\sqrt{10}
\,
d_{KMN}\,{F}^K \wedge  {F}^M \wedge  {F}^N
\,,
\label{eq:CS}
\end{equation}
with the cubic symmetric ${\rm E}_{6(6)}$-invariant tensor $d_{KMN}$.
The Lagrangian (\ref{eq:D5}) is manifestly invariant under global $\mathfrak{e}_{6(6)}={\rm Lie}\,{\rm E}_{6(6)}$ transformations
\begin{equation}
\delta_\Lambda A_\mu{}^M = -\Lambda_N{}^M\,A_\mu{}^N\,,\quad
\delta_\Lambda {M}_{MN} =\Lambda_{M}{}^{K} {M}_{NK}+ \Lambda_{N}{}^{K} {M}_{MK}
\,,
\label{eq:E6}
\end{equation}
with the matrices $\Lambda_M{}^N$ in the fundamental representation ${\cal R}_1$ of $\mathfrak{e}_{6(6)}$.
I.e., the 27 vector fields transform in the representation ${\cal R}_1$, while the 42 scalar fields parametrizing ${M}_{MN}$ transform in the non-linear representation realized on the coordinates of the coset space ${\rm E}_{6(6)}/{\rm USp}(8)$.
In turn, the Lagrangian (\ref{eq:D5})  combines the most general two-derivative couplings for the above field content, invariant under the bosonic symmetry (\ref{eq:E6}). The most straightforward construction of maximal $D=5$ supergravity starts from the bosonic Lagrangian (\ref{eq:E6}) and determines the couplings of the fermionic sector by imposing maximal supersymmetry. In particular, this uniquely fixes the a priori unrelated relative factors between the different terms in the bosonic Lagrangian.

Maximal $D=5$ supergravity can be obtained by toroidal reduction from $D=11$ supergravity \cite{Cremmer:1978km} and IIB supergravity \cite{Schwarz:1983wa,Howe:1983sra}, respectively. This corresponds to a Kaluza-Klein split of the the coordinates of the higher-dimensional space-time into
\begin{equation}
\{ X^{\hat\mu} \} \longrightarrow \left\{
\begin{array}{rll}
\{x^\mu, y^a\} & a = 1, \dots, 6\quad&11{\rm D} \\[.5ex]
\{x^\mu, y^i\} & i = 1, \dots, 5\quad &{\rm IIB} 
\end{array}
\right.
,
\label{eq:KKsplit}
\end{equation}
where $\mu=0 \dots, 4$, labels the $D=5$ coordinates, 
and $y^a$ and $y^i$ are the coordinates of the internal six-torus $T^6$ and five-torus $T^5$, respectively.
Toroidal reduction imposes the independence of all fields from the internal torus coordinates. Equivalently, it can be understood as a Fourier expansion of the higher-dimensional fields on the torus, and truncating the theory to the singlets under the ${\rm U}(1)$ isometries of the torus.
It requires some work to reorganize the different higher-dimensional fields appearing in the reduction into the ${\rm E}_{6(6)}$-covariant objects that build the Lagrangian (\ref{eq:D5}). E.g.\ the 27 vectors of the $D=5$ theory, combine fields of various higher-dimensional origin according to
\begin{align}
11{\rm D} \,:\;\;& A_\mu{}^M\;=\; \{ A_{\mu}{}^{a}, A_{\mu\,ab}, , A_{\mu\,abcde} \}\,,
\nonumber\\
{\rm IIB}\,:\;\;& A_\mu{}^M \;=\;  \{ A_{\mu}{}^{i}, A_{\mu\,i\alpha}, A_{\mu\,ijk}, A_{\mu\,ijklm\alpha} \}\,.
\label{eq:A-origin}
\end{align}
The index $\alpha$ here denotes a doublet under the global ${\rm SL}(2)$ symmetry of IIB supergravity. The decomposition (\ref{eq:A-origin}) thus exhibits components from the D=11 metric, three-form and six-form for the first line, and IIB metric, two-forms, four-form, and six-forms, for the second line, respectively. 
The precise embedding of the five-dimensional fields into the higher-dimensional fields is done in the flat basis, i.e.\ after flattening and unflattening the $p$-form indices with the respective vielbeins.
The decompositions (\ref{eq:A-origin}) follow the different breakings of the exceptional symmetries group
\begin{align}
11{\rm D} \,:\;\;& {\rm E}_{6(6)} \longrightarrow \mathbb{R}_{11{\rm D}} \times {\rm SL}(6)\,,
\nonumber\\
{\rm IIB}\,:\;\;& {\rm E}_{6(6)} \longrightarrow \mathbb{R}_{\rm IIB} \times {\rm SL}(5) \times {\rm SL}(2)  \,,
\label{eq:E-origin}
\end{align}
into the subgroups that have manifest geometric higher-dimensional origin descending from internal diffeomorphisms.
For the $D=5$ scalar fields with their non-linear couplings, the embedding into higher dimensions is more intricate but conveniently organized by
the algebraic structure of the embeddings (\ref{eq:E-origin}). 
Specifically, the scalar matrix is built as ${M}_{MN}=({V}{V}^\top)_{MN}$ from a coset representative ${V}\in {\rm E}_{6(6)}$ in suitable triangular gauge \cite{Cremmer:1997ct}. E.g.\ for the IIB embedding, the explicit embedding reads
\begin{equation}
{ V} = 
{\rm exp}\left[\varepsilon^{ijklm}\,A_{ijkl} \, t_{(+6)\,m}\right] \,
{\rm exp}\left[A_{ij}{}^\alpha\,t_{(+3)}{}_{\alpha}^{ij}\right]\,
{\cal V}_5\,{\cal V}_2\,
{\rm exp} \left[\Phi\, t_{(0)}\right]
\;.
\label{eq:V27B}
\end{equation}
Here, $t_{(0)}$ is the E$_{6(6)}$ generator associated to the $\mathbb{R}_{\rm IIB} $ grading of (\ref{eq:E-origin}), and the 
$t_{(+n)}$ refer to the E$_{6(6)}$ generators of positive grading under $t_{(0)}$.
The decomposition (\ref{eq:V27B}) shows the scalar components of the IIB two-forms and five-form, respectively,
together with the matrices ${\cal V}_2\in{\rm SL}(2)$, describing the IIB dilaton/axion sector, and the matrix 
${\cal V}_5\in{\rm SL}(5)$, which captures the internal block of the ten-dimensional metric as $G_5=e^{-\Phi/2}\,{\cal V}_5({\cal V}_5)^\top$\,.
The  resulting matrix ${M}_{MN}$ decomposes according to the split (\ref{eq:A-origin})
\begin{eqnarray}
{M}_{MN} &=& \left(
\begin{array}{cccc}
{M}_{i,m}&{M}_i{}^{m\beta}&{M}_{i}{}^{mnp}&{M}_{i}{}^{\beta}\\
{M}^{i\alpha}{}_{m}&{M}^{i\alpha,}{}^{m\beta}&{M}^{i\alpha,mnp}&{M}^{i\alpha,}{}^{\beta}\\
{M}^{ijk}{}_{m}&{M}{}^{ijk,m\beta}&{M}^{ijk,mnp}&{M}^{ijk,\beta}\\
{M}^\alpha{}_{m}&{M}^{\alpha,m\beta}&{M}^{\alpha,mnp}&{M}^{\alpha,\beta}
\end{array}
\right)
\;,
\label{eq:M2B}
\end{eqnarray}
where the explicit expressions for the various blocks are obtained from evaluating the exponential in (\ref{eq:V27B}).
We refer to \cite{Baguet:2015xha} for the explicit formulas and just give two examples
\begin{eqnarray}
A_{mn}{}^\alpha
 &=&
\frac16\, \sqrt{2}\,  ({\rm det} \,G_5)^{2/3}\, \varepsilon^{\alpha\beta}\,\varepsilon_{mnpqr} m_{\beta\gamma} \,{M}^{\gamma,pqr}{}  
   \,,\nonumber\\
A_{klmn} &=&{\frac18}\, ({\rm det} \,G_5)^{2/3}\,\varepsilon_{klmnp}\,m_{\alpha\beta} \,{M}^{\alpha,p\beta}
 \,,
 \label{eq:embedding_scalars}
 \end{eqnarray}
for extracting the internal components of the IIB $p$-forms from ({\ref{eq:M2B}). Here, $\varepsilon^{\alpha\beta}$ and  $\varepsilon_{mnpqr}$ are the constant
fully antisymmetric tensors of ${\rm SL}(2)$ and ${\rm SL}(5)$, respectively, and $m_{\alpha\beta}$ is the IIB dilaton/axion matrix $m={\cal V}_2({\cal V}_2)^\top$, which in turn is extracted from (\ref{eq:M2B}) as $m^{\alpha\beta}=({\rm det} \,G_5)^{2/3}\,{M}^{\alpha,\beta}$.

To summarize, after toroidal reduction of IIB supergravity on a five-torus (or equivalently $D=11$ supergravity on a six-torus), and subsequent reorganization of the fields, the bosonic sector of the resulting five-dimensional theory takes the form (\ref{eq:D5}), manifestly invariant under the global symmetry group ${\rm E}_{6(6)}$. The action of this group can be used as a solution generating symmetry for ten-dimensional IIB solutions with five commuting isometries.
In practice, the reorganization of the higher-dimensional fields into E$_{6(6)}$-covariant objects  is fairly involved, especially in the fermionic sector. On the other hand, the premise of E$_{6(6)}$ as the global symmetry in $D=5$ dimensions, determines the theory up to a few coefficients which are then fixed by supersymmetry.

\section{Gauged maximal supergravity: the embedding tensor}
\label{sec:gauged}

Maximal supergravities obtained by torus reduction from ten and eleven dimensions are abelian gauge theories in which none of the matter fields are charged under the local ${\rm U}(1)$ gauge symmetries associated with the vector fields.
Non-abelian gauge symmetry can be introduced upon promoting a subgroup ${\rm G}_{0}$ of the global exceptional symmetry group ${\rm E}_{d(d)}$ to a local gauge symmetry.  This was first achieved in \cite{deWit:1982bul} for the ${\rm SO}(8)$ gauged supergravity in $D=4$ dimensions, and generalized in \cite{Pernici:1984xx,Gunaydin:1985cu} to higher dimensions. Every such gauging breaks the exceptional symmetry group down to the local gauge group, yet, the exceptional symmetry is fundamental to constructing and classifying the gauged theories. In particular, the fact that the vector fields in the ungauged theory transform in a fundamental representation ${\cal R}_1$ of the exceptional group (rather than in the adjoint one), puts severe constraints on the possible choices of gauge groups  ${\rm G}_{0}\subset{\rm E}_{d(d)}$. This is most conveniently described by the embedding tensor formalism \cite{Nicolai:2000sc,deWit:2002vt,deWit:2004nw}. 

Here, we briefly review the resulting structure for $D=5$ maximal gauged theories, based on the exceptional group  ${\rm E}_{6(6)}$, following \cite{deWit:2004nw}. The global symmetry  action (\ref{eq:E6}) of the ungauged theory can be expanded as
\begin{eqnarray}
\delta_\Lambda &=& \Lambda^\alpha \mathbb{T}_\alpha \,\cdot
\;,
\label{eq:action}
\end{eqnarray}
into the action of the 78 generators $\mathbb{T}_\alpha$ of $\mathfrak{e}_{6(6)}={\rm Lie}\, E_{6(6)}$,
closing into the algebra
\begin{eqnarray}
[\,\mathbb{T}_\alpha , \, \mathbb{T}_\beta\,] &=& f_{\alpha\beta}{}^\gamma\,\mathbb{T}_\gamma
\;.
\label{eq:E6alg}
\end{eqnarray}
A gauging of the theory is characterized by a choice of 27 (not necessarily independent) generators $X_M \in \mathfrak{e}_{6(6)}$ whose action is promoted into a local symmetry upon the introduction of covariant derivatives
\begin{equation}
D_\mu =\partial_\mu - g A_\mu{}^M X_M
\;,
\label{eq:cov}
\end{equation}
with a gauge coupling constant $g$. The choice of generators is encoded by a constant embedding tensor $\Theta_M{}^\alpha$ as
\begin{eqnarray}
X_M &\equiv & \Theta_M{}^\alpha\,\mathbb{T}_\alpha
\;.
\label{eq:Theta}
\end{eqnarray}
Not every choice of generators (\ref{eq:Theta}) leads to a consistent theory.
As an obvious constraint, the generators $X_M$ must close into a subalgebra. In addition, the adjoint representation 
of this algebra must be embedded in the fundamental representation ${\cal R}_1$.
Moreover, it is only for very particular subalgebras that the deformation is compatible with maximal supersymmetry.
Fortunately, all these consistency conditions can be encoded in a simple set
of algebraic equations for the embedding tensor~$\Theta_M{}^\alpha$, which we can spell out explicitly as
\begin{align}
3\,(\mathbb{T}_\beta \mathbb{T}_\alpha)_M{}^N\,\Theta_N{}^\beta+2\,\Theta_M{}^\alpha=\,& 0
\,,\label{eq:con1}\\
\Theta_{ P}{}^\beta
    \Theta_{ N}{}^\alpha  (\mathbb{T}_\beta)_{M}{}^{ N}+
    \Theta_{ P}{}^\beta 
    \Theta_{ M}{}^\gamma f_{\beta\gamma}{}^\alpha  =\,& 0
\;,
\label{eq:con2}
\end{align}
in terms of the generators and structure constants of $\mathfrak{e}_{6(6)}$.
In terms of representations, the linear constraint (\ref{eq:con1}) states that the embedding tensor 
is entirely contained in the ${\bf 351}$ representation within the full tensor product of its indices
\begin{equation}
{\bf 27} \otimes {\bf 78} = {\bf 27}\oplus{\bf 351} \oplus {\bf 1728}'
\,.
\end{equation}
This condition follows from compatibility of the deformation with maximal supersymmetry.
The quadratic constraint (\ref{eq:con2}) states that the ${\bf 27}'\oplus{\bf 1728}$ is absent in the
tensor product $({\bf 351}\otimes{\bf 351})_{\rm sym}$ of two embedding tensors. It can be rewritten in the form
\begin{equation}
[X_M, X_N] = -X_{MN}{}^K\,X_K
\,,
\label{eq:gg}
\end{equation}
which shows the closure of the algebra of generators (\ref{eq:Theta}), with the structure constants
of the gauge algebra given by $X_{MN}{}^K = \Theta_M{}^\alpha\,(\mathbb{T}_\alpha)_N{}^K$\,.
In turn, one can show that every solution to the system of algebraic equations (\ref{eq:con1}), (\ref{eq:con2})
defines a consistent theory with maximal local supersymmetry
and gauge algebra spanned by the generators~(\ref{eq:Theta}).
The exceptional symmetry structure of the ungauged theory thus allows a concise classification of 
its consistent non-abelian deformations. In the following, we will briefly sketch the structure of the resulting theories.

The bosonic Lagrangian of $D=5$ maximal gauged supergravity  is given by a deformation of (\ref{eq:D5}) to
\begin{align}
 {\cal L}_{\rm gauged} =\,& \sqrt{-{\rm g}}\,\Big(R
 +\tfrac{1}{24} D_{\mu}{M}^{MN} D^{\mu}{M}_{MN}
-\tfrac{1}{4}{M}_{MN}{F}^{\mu\nu M}{F}_{\mu\nu}{}^N-V_{\rm pot}\Big)
\nonumber\\
&{}
+{\cal L}_{\rm top} \,,
\label{eq:D5gauged}
\end{align}
whose different terms and ingredients we will now go through. 
Derivatives on the scalar fields are covariantized according to (\ref{eq:E6}), (\ref{eq:cov}), and the formely abelian field strengths
have been replaced by their non-abelian counterparts
\begin{eqnarray}
  {F}_{\mu\nu}{}^{M} &=& \partial_\mu A_\nu{}^{M}
  -\partial_\nu
  A_\mu{}^{M} +  g\,X_{[{NP}]}{}^{M}
  \,A_\mu^{N} A_\nu{}^{P} 
  +g\,Z^{MN}B_{\mu\nu\,N}
  \,.
\label{eq:defF}
\end{eqnarray}
As a new feature, compared to standard Yang-Mills theories, these non-abelian field strengths carry a St\"uckelberg-type coupling to a set of 27 two-form gauge potentials $B_{\mu\nu\,M}$, triggered by the embedding tensor, via the antisymmetric matrix 
\begin{equation}
Z^{MN}=d^{MKL}\, \Theta_K{}^\alpha(t_\alpha)_L{}^N=-Z^{NM}\,.
\label{eq:Z}
\end{equation} 
The presence of the two-form fields in (\ref{eq:defF}) is required in order to ensure covariance of the field strengths under non-abelian gauge transformations which read
\begin{align}
\delta A_\mu{}^{M} =\,&  D_\mu\Lambda^{M} -
g\,Z^{MN}{}\,\Xi_{\mu \, N} \,,
\nonumber\\
\delta B_{\mu\nu\,M} =\,&
    2\,D_{[\mu}\Xi_{\nu]\,M} +2\,d_{MKL}
  \left(\Lambda^{K} {F}_{\mu\nu}{}^{{L}}
  -  A_{[\mu}{}^{{K}}
  \delta A_{\nu]}{}^{{L}} \right)
  \,.
  \label{eq:gaugeAB}
\end{align}
These entangled gauge transformations emerge as the beginning of the non-abelian tensor hierarchy \cite{deWit:2005hv,deWit:2008ta,Palmkvist:2013vya,Cederwall:2019qnw},
which extends to all the higher-rank $p$-forms.
The linear constraint (\ref{eq:con1}) proves to be crucial for consistency of the system. Even though this constraint initially arises from compatibility with maximal supersymmetry, its necessity can  already be recognized within this purely bosonic structure. This provides a glimpse of the more general structures which we will encounter in the next section.

The two-form potentials also enter the non-abelian topological term ${\cal L}_{\rm top}$ in (\ref{eq:D5gauged}) which is now defined by
\begin{equation}
d{\cal L}_{\rm top}
= \tfrac{1}{6}\sqrt{10}
\,\big(
d_{MNK}\,{F}^M \wedge  {F}^N \wedge  {F}^K
+g\,Z^{MN} H_M \wedge dH_N\big)
\,,
\label{eq:CSgauged}
\end{equation}
where $H_M$ is the covariant field strength of the two-form gauge potentials
\begin{equation}
{H}_{\mu\nu\rho\,M} = 
3D_{[\mu}  B_{\nu\rho]\,M} 
+6d_{MPQ} \,A_{[\mu}{}^P  \partial_{\nu} A_{\rho]}{}^Q 
+2g\, d_{MPQ} \, X_{RS}{}^Q A_{[\mu}{}^P A_\nu{}^RA_{\rho]}{}^S 
\,,
\label{eq:defH}
\end{equation}
up to terms that vanish under the projection with $Z^{MN}$.
The presence of the two-form gauge potentials in the non-abelian Lagrangian (\ref{eq:D5gauged}) gives rise to additional field equations, which precisely yield the first-order duality equations, relating vector and tensor fields as
\begin{equation}
Z^{KM}\,M_{MN}\,F_{\mu\nu}{}^N 
 =\frac{\sqrt{10}}{6}\, \sqrt{-{\rm g}}\, \varepsilon_{\mu\nu\rho\sigma\tau}\,
  H^{\rho\sigma\tau}{}_{M} Z^{MK}\,.  
\label{eq:duality}
\end{equation}
This is an extension of the standard on-shell duality between abelian $p$-forms and $(D-p-2)$-forms to the non-abelian case, consistent with the gauge transformations (\ref{eq:gaugeAB}).
Even though the Lagrangian of the gauged theory (\ref{eq:D5gauged}) carries more fields than its ungauged limit (\ref{eq:D5}), the total number of propagating degrees of freedom is thus conserved, since (\ref{eq:duality}) ensures that the new fields do not introduce additional degrees of freedom.

For a given gauging, i.e.\ a fixed choice of the embedding tensor, the gauge symmetries (\ref{eq:gaugeAB}) associated with the tensor fields, can be used to eliminate a subset of the vector fields, on which they act as a shift symmetry. As a consequence, the corresponding tensor fields turn into topologically massive tensor fields~\cite{Townsend:1983xs}, with equation (\ref{eq:duality}) turning into a first-order equation of the type
\begin{equation}
 H_{\mu\nu\rho\,M}= -\frac{g}{2\,\sqrt{10}}\,M_{MN} Z^{NK}\,\sqrt{-{\rm g}}\,\varepsilon_{\mu\nu\rho\sigma\tau}\,B^{\sigma\tau}{}_{K} 
 \,,  
 \label{eq:D5HB}
\end{equation}
for the relevant two-forms.
For instance, for the maximal supergravity with gauge group ${\rm SO}(6)$, 15 out of the 27 vector fields correspond to the non-abelian gauge vectors of ${\rm SO}(6)$, invariant under the shift symmetry $\delta_\Xi$, while the 12 remaining vector fields (transforming in a non-trivial representation of ${\rm SO}(6)$) can be eliminated by this shift symmetry. After this gauge fixing, the theory carries 15 gauge vectors and 12 topologically massive tensor fields, thus preserving the total number of degrees of freedom. It is in this gauge-fixed form, that the ${\rm SO}(6)$ gauged theory was first constructed in~\cite{Gunaydin:1985cu}. For other gauge groups the decomposition of the 27 degrees of freedom into massless gauge vectors and topologically massive tensor fields may be different. The virtue of the above formulation is the covariant expression for all couplings in which the distribution of the degrees of freedom for any particular theory is automatically selected by evaluating the expressions for its specific embedding tensor.

Let us finally give the form of the scalar potential that appears in the gauged theory (\ref{eq:D5gauged}) which 
takes the form
\begin{equation}
V_{\rm pot} =
\frac{g^2}{30}\,M^{MN}X_{MP}{}^Q
\left(5\, X_{NQ}{}^P + X_{NR}{}^S \,M^{PR}M_{QS}
\right)
,
\label{eq:potentialN8}
\end{equation}
bilinear in the embedding tensor through the structure constants of the gauge group (\ref{eq:gg}). The scalar dependence
arises through the group matrix $M_{MN}$. Expressed in terms of the structure constants, the potential is manifestly gauge invariant.
However, unlike the rest of the bosonic theory (\ref{eq:D5gauged}), 
whose structure is essentially determined by consistency with the non-abelian gauge symmetries,
the precise form of (\ref{eq:potentialN8}), and in particular the relative coefficient between its terms follows from 
consistency with supersymmetry.

\begin{figure}[tb]
\center
\includegraphics[scale=.52]{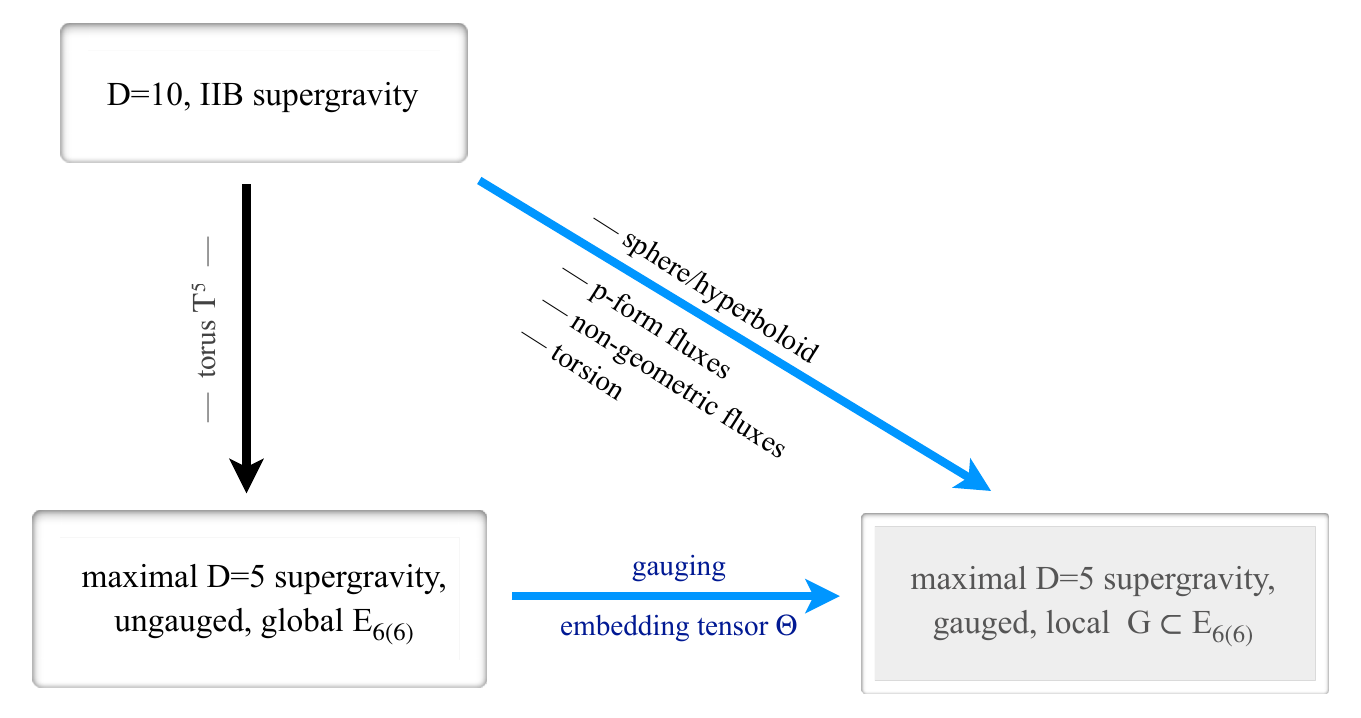}
\caption{
Gauging supergravity}
\label{fig:gauged}
\end{figure}

The universal form of the Lagrangian of the gauged theory 
is of particular importance for the study of the actions associated to
specific flux compactifications of higher-dimensional theories.
E.g.\ in flux compactifications of higher-dimensional supergravity on (twisted) tori,
the flux parameters show up as deformation parameters among the components
of the embedding tensor and can be uniquely identified by purely
group-theoretical means, see~\cite{Samtleben:2008pe,Trigiante:2016mnt} and references therein.
The general expression~(\ref{eq:potentialN8}) then allows to read off the full scalar
potential of the effective theory upon specifying the particular choice of~$\Theta_M{}^\alpha$,
as schematically depicted in Figure~\ref{fig:gauged}.

Let us also note that by construction two gaugings whose embedding tensors are related by a global ${\rm E}_{6(6)}$ transformation are equivalent upon field redefinition (\ref{eq:E6}). In particular, such a redefinition relocates the scalar origin $M_{MN}=\delta_{MN}$. The set of inequivalent gaugings thus corresponds to the ${\rm E}_{6(6)}$ orbits of solutions to (\ref{eq:con1}), (\ref{eq:con2}).

\section{Exceptional field theory and consistent truncations}
\label{sec:exft}

We have reviewed above the role of the exceptional symmetry group E$_{6(6)}$ in the formulation of the $D=5$ ungauged maximal supergravity, as well as its role in the construction and classification of its deformations, the gauged maximal supergravities. The universal structure of all $D=5$ maximal supergravities, together with the fact that many of these theories can be obtained by reduction from higher dimensions, invites the question of to what extent the exceptional structure can already be realized in higher dimensions. This is achieved by exceptional field theory (ExFT), the duality covariant formulation of $D=11$ and IIB supergravity \cite{Hohm:2013pua}. Here, we will  describe the corresponding exceptional field theory, associated with the exceptional group  E$_{6(6)}$, following \cite{Hohm:2013vpa}.

Exceptional field theory is based on a Kaluza-Klein split (\ref{eq:KKsplit}) of the coordinates of the higher-dimensional spacetime, into an external, $D=5$, space-time, and an internal spacetime. However, unlike in the torus reduction, discussed above, the fields maintain their dependence on all, external and internal, coordinates. The Lagrangian of E$_{6(6)}$ exceptional field theory is modelled after the Lagrangians (\ref{eq:D5}) and (\ref{eq:D5gauged}) of $D=5$ maximal supergravity.
In principle, this Lagrangian could be obtained starting from the split (\ref{eq:KKsplit}) and 
reorganizing the higher-dimensional fields into ${\rm E}_{6(6)}$-covariant objects as in (\ref{eq:A-origin}), (\ref{eq:M2B}), albeit in presence of the internal coordinates. However, as for the maximal $D=5$ supergravities described above, the more straightforward approach to their construction is to first identify the relevant symmetries and to proceed to directly formulate the most general invariant two-derivative action in terms of appropriate covariant objects.

For exceptional field theory, the relevant symmetry structure is given by the generalized internal diffeomorphisms \cite{Coimbra:2011ky,Berman:2012vc}. These transformations unify internal diffeomorphisms and internal gauge symmetries into a single ${\rm E}_{6(6)}$-covariant structure  with symmetry parameter $\Lambda^M$ in the fundamental representation of ${\rm E}_{6(6)}$. Explicitly, their action on a generalized one-form $V_M$ is defined as 
\begin{equation}
{\cal L}_\Lambda V_M  = \Lambda^N\partial_N V_M + 6\left[\partial_M \Lambda^N\right]_{\rm adj} \,V_N
\,.
\label{eq:genL}
\end{equation}
Compared to standard diffeomorphisms, the second term of the transformation (\ref{eq:genL}) carries the explicit projector
\begin{align}
6\left[\partial_M \Lambda^N\right]_{\rm adj}
=\,& 6\,(\mathbb{T}^\alpha)_M{}^N (\mathbb{T}_\alpha)_K{}^L\,\partial_L\Lambda^K
\nonumber\\
=\,&    \partial_M\Lambda^N  +\tfrac1{3}\, \partial_K\Lambda^K\,\delta_N{}^M
-10\,d^{NLR}d_{MKR}\, \partial_L\Lambda^K\,, 
\end{align}
onto the adjoint representation of ${\rm E}_{6(6)}$.
The projection guarantees a consistent action of generalized diffeomorphisms on group-valued matrices, such as $M_{MN}$. Likewise, it ensures compatibility with the ${\rm E}_{6(6)}$ invariant tensors, such as
\begin{equation}
{\cal L}_\Lambda\left( V_KW_MU_N\,d^{KMN}\right) = 0
\,,\quad \mbox{etc.}\,.
\end{equation}
The factor 6 in front of the second term of (\ref{eq:genL}) is determined by closure of the algebra.
Finally, in (\ref{eq:genL}), the internal coordinates are formally embedded into the 
27-dimensional representation of ${\rm E}_{6(6)}$ as
\begin{equation}
\partial_i \hookrightarrow \partial_M
\,.
\label{eq:incl}
\end{equation}
Closure of the algebra of transformations (\ref{eq:genL}) further requires the so-called section constraint, expressed as an ${\rm E}_{6(6)}$-covariant condition bilinear in the internal derivatives
\begin{equation}
d^{KMN}\,\partial_M \Phi_1 \partial_N \Phi_2  = 0
\,,
\label{eq:section}
\end{equation}
acting on any couple $\{\Phi_1, \Phi_2\}$ of fields or gauge parameters.

${\rm E}_{6(6)}$ exceptional field theory can now be constructed as a two-derivative action invariant under the generalized diffeomorphisms (\ref{eq:genL}).
Its bosonic field content is identical with the field content of five-dimensional maximal supergravity (\ref{eq:D5gauged}), i.e.\ built from the fields
\begin{equation}
\left\{
g_{\mu\nu}, {\cal A}_{\mu}{}^M, {\cal B}_{\mu\nu\,M}, {\cal M}_{MN} \right\}
\,,\quad
\mu, \nu =0, \dots, 4\;,\quad
M=1, \dots, 27\,,
\label{eq:ExFTfieldsE6}
\end{equation}
and combines the $5\times 5$ external metric $g_{\mu\nu}$ with 
27 vectors, two-forms, and the scalar matrix ${\cal M}_{MN}$ which parametrizes the coset space ${\rm E}_{6(6)}/{\rm USp}(8)$\,. In contrast to (\ref{eq:D5gauged}) however, all fields still depend on all, external and internal, coordinates, the latter dependence only constrained by the section constraint (\ref{eq:section}). We use calligraphic letters for the ExFT fields content (\ref{eq:ExFTfieldsE6}) to mark the difference with the analogous objects in the $D=5$ theories.
The complete bosonic Lagrangian of ${\rm E}_{6(6)}$ exceptional field theory formally ressembles the $D=5$ theory (\ref{eq:D5gauged}) and reads \cite{Hohm:2013vpa}

\begin{align}
 {\cal L}_{{\rm ExFT}} =\,& \sqrt{-g}\,\Big( \widehat{{\cal R}}
 +\tfrac{1}{24} {\cal D}_{\mu}{\cal M}^{MN} {\cal D}^{\mu}{\cal M}_{MN}
-\tfrac{1}{4}{\cal M}_{MN}{\cal F}^{\mu\nu M}{\cal F}_{\mu\nu}{}^N-V\Big)
\nonumber\\
&{}
+{\cal L}_{\rm top} \,.
\label{eq:D5ExFT}
\end{align}
Here, the Einstein-Hilbert term is built from the modified Ricci scalar $ \widehat{{\cal R}}$, constructed
from the external metric $g_{\mu\nu}$, upon covariantizing external derivatives as
\begin{equation}
\partial_\mu g_{\nu\rho} \rightarrow 
(\partial_\mu -{\cal L}_{{\cal A}_\mu})\,g_{\nu\rho} \equiv
\partial_\mu - {\cal A}_\mu{}^K \partial_K g_{\nu\rho}-\frac23\,\partial_K{\cal A}_\mu{}^Kg_{\nu\rho}
\,,
\end{equation}
according to the action of internal diffeomorphisms onto the external metric.
The covariant derivatives on the scalar matrix are defined by
\begin{equation}
{\cal D}_{\mu}{\cal M}_{MN} =
(\partial_\mu -{\cal L}_{{\cal A}_\mu})\,{\cal M}_{MN}
\,,
\label{eq:covDE6}
\end{equation}
with a connection defined via the action (\ref{eq:genL}) of generalized diffeomorphisms.
The non-abelian field strengths in (\ref{eq:D5ExFT}) are defined by
\begin{align}
{\cal F}_{\mu\nu}{}^N  = \,&  2\, \partial_{[\mu} {\cal A}_{\nu]}{}^N 
-2\,{\cal A}_{[\mu}{}^K \partial_K {\cal A}_{\nu]}{}^N 
+10\, d^{NKR}d_{PLR}\,{\cal A}_{[\mu}{}^P\,\partial_K {\cal A}_{\nu]}{}^L
\nonumber\\
&{}
+ 10 \, d^{NKL}\,\partial_K {\cal B}_{\mu\nu\,L}
\,.
\label{eq:FExFT}
\end{align}
The term quadratic in the vector fields is the standard Yang-Mills contribution for the non-abelian gauge symmetry 
of (\ref{eq:genL}), obtained from the commutator of covariant derivatives (\ref{eq:covDE6}). The coupling to the two-form fields ${\cal B}_{\mu\nu\,M}$ is reminiscent of the structure (\ref{eq:defF}) and required in order to ensure covariance of these field strengths under the full non-abelian gauge transformations which read
 \begin{align}
   \delta {\cal A}_\mu{}^M \ &= \ {\cal D}_\mu \Lambda^M  
   -10\, d^{MNK} \partial_K\Xi_{\mu N}\,, \nonumber\\
   \delta {\cal B}_{\mu\nu M} \ &= \ 2\,{\cal D}_{[\mu}\Xi_{\nu]\,M} 
   +d_{MKL} \left(\Lambda^K{\cal F}_{\mu\nu}{}^{L}
   - {\cal A}_{[\mu}{}^K\, \delta {\cal A}_{\nu]}{}^L\right).
 \label{deltaAB}
 \end{align}
They constitute the beginning of the non-abelian tensor hierarchy in exceptional field theory.
The topological term ${\cal L}_{\rm top} $ in (\ref{eq:D5ExFT}) is defined via
\begin{equation}
d{\cal L}_{\rm top}
= \tfrac{1}{6}\sqrt{10}
\left(
d_{MNK}\,{\cal F}^M \wedge  {\cal F}^N \wedge  {\cal F}^K
-40\, d^{MNK}{\cal H}_M\,  \wedge \partial_N{\cal H}_K
\right)
\,,
\label{CSlike}
\end{equation}
with the non-abelian field strength of the two-form fields given by
\begin{align}
 {\cal H}_{\mu\nu\rho\,M} =\,&
3\,{D}_{[\mu} {\cal B}_{\nu\rho]\,M}
-3\,d_{MKL}\, {\cal A}_{[\mu}{}^K\,\partial_{\vphantom{[}\nu} {\cal A}_{\rho]}{}^L 
+ 2\,d_{MKL}\, {\cal A}_{[\mu}{}^K {\cal A}_{\vphantom{[}\nu}{}^P \partial_P {\cal A}_{\rho]}{}^L 
\nonumber\\
&{}
-10\, d_{MKL}d^{LPR}d_{RNQ}\, {\cal A}_{[\mu}{}^K {\cal A}_{\vphantom{[}\nu}{}^N\,\partial_P {\cal A}_{\rho]}{}^Q
\,,
\label{eq:HExFT}
  \end{align}
up to terms that vanish under the projection with $d^{MNK} \partial_K$. The field strengths of vector and tensor fields are related by a generalized Bianchi identity
\begin{equation}
3\,{\cal D}_{[\mu} {\cal F}_{\nu\rho]}{}^N  =  {10} \, d^{NKL}\,\partial_K {\cal H}_{\mu\nu\rho\,L}
\,,
\end{equation}
as well as by the tensor field equations
\begin{equation}
d^{MNK}\partial_K  \left(\sqrt{-{g}}\,{\cal M}_{NL} {\cal F}^{\mu\nu L}
 +\tfrac16 \sqrt{10}\,  \varepsilon^{\mu\nu\rho\sigma\tau}\,
  {\cal H}_{\rho\sigma\tau N}\right) = 0
  \,,
\label{eq:ExFTFH}
 \end{equation}
reminiscent of the duality equations in $D=5$ gauged supergravity (\ref{eq:duality}).
The latter equations ensure, that the two-form tensor fields do not constitute independent degrees
of freedom in the Lagrangian (\ref{eq:D5ExFT}).
Finally, the so-called potential term $V$ in (\ref{eq:D5ExFT}) is built from bilinears in internal derivatives and reads
\begin{align}
  V \ = \ &-\frac{1}{24}\,{\cal M}^{MN}\partial_M{\cal M}^{KL}\,\partial_N{\cal M}_{KL}+\frac{1}{2}\, {\cal M}^{MN}\partial_M{\cal M}^{KL}\partial_L{\cal M}_{NK}
 \label{eq:potExFT}\\
  &-\frac{1}{2}\,g^{-1}\partial_Mg\,\partial_N{\cal M}^{MN}
  -\frac{1}{4} \, {\cal M}^{MN}\left(
  g^{-1}\partial_Mg\,g^{-1}\partial_Ng
 +\partial_Mg^{\mu\nu}\partial_N g_{\mu\nu}\right)
 \,. 
  \nonumber
 \end{align}
 Although not manifest, it is straightforward to show that the potential term $\sqrt{-g}\,V$ 
 is invariant under generalized diffeomorphisms (\ref{eq:genL}), up to total derivatives.
 This term can be understood more geometrically, in terms of a generalized Ricci scalar 
of the internal geometry that, however, also depends on the external metric \cite{Coimbra:2012af,Godazgar:2014nqa}.
 
 All other terms in the Lagrangian (\ref{eq:D5ExFT})  are explicitly built from covariant objects, thus manifestly 
 invariant under the generalized diffeomorphisms. The Lagrangian combines the most general set of gauge-invariant two-derivative terms. As it turns out, also the relative coefficients of this Lagrangian are uniquely fixed by further requiring invariance under properly modified external diffeomorphisms, given by
 \begin{align}
 \delta e_{\mu}{}^{a} =\,\,& \xi^{\nu}{\cal D}_{\nu}e_{\mu}{}^{a}
 + {\cal D}_{\mu}\xi^{\nu} e_{\nu}{}^{a}\;, \nonumber\\
\delta {\cal M}_{MN} =\,\,& \xi^\mu \,{\cal D}_\mu {\cal M}_{MN}\;,\nonumber\\
\delta {\cal A}_{\mu}{}^M =\,\,& \xi^\nu{\cal F}_{\nu\mu}{}^M + {\cal M}^{MN}g_{\mu\nu} \,\partial_N \xi^\nu
\;,\nonumber\\
\delta {\cal B}_{\mu\nu\,M} =\,\,& \tfrac{1}{2\sqrt{10}}\,\xi^\rho\,
 \epsilon_{\mu\nu\rho\sigma\tau}\, {\cal F}^{\sigma\tau\,N} {\cal M}_{MN} 
 -d_{MKL}\,{\cal A}_{[\mu}{}^K\, \delta {\cal A}_{\vphantom{\mu}\nu]}{}^L
 \,. 
 \label{eq:skewD}
\end{align}
For a diffeomorphism parameter with $\partial_M\xi^\mu=0$, these transformations reduce to the standard (covariantized) action of $D=5$ diffeomorphisms.

In summary, the Lagrangian (\ref{eq:D5ExFT}) is the unique two-derivative action invariant under generalized internal and external diffeomorphisms (\ref{eq:genL}) and (\ref{eq:skewD}), provided all fields satisfy the section constraint (\ref{eq:section}).
 The latter severely restricts the dependence of the fields on the internal coordinates. 
 It can be systematically solved by describing the embedding of the internal coordinates (\ref{eq:incl}) as
 \begin{equation}
 \partial_M = {\cal E}_M{}^m\,\partial_m\,,
\end{equation}
 in terms of a constant section matrix ${\cal E}_M{}^m$ whose different choices are classified by the inequivalent maximal embeddings of the linear groups ${\rm GL}(d)$ into the exceptional group, see \cite{Inverso:2017lrz,Cederwall:2017fjm} for a systematic discussion. For E$_{6(6)}$, the relevant embeddings correspond to the decompositions (\ref{eq:E-origin}). Breaking 
\begin{align}
\rm{E}_{6(6)} \supset~&    \mathbb{R}_{11{\rm D}} \times {\rm SL}(6)
\,,\nonumber\\
{\bf 27} \longrightarrow ~&  6_{+1} \oplus 15'_{0} \oplus 6_{-1}
\,, \nonumber\\
\partial_M \longrightarrow ~& \{ \partial_{a} , \partial^{ab} , \partial^{abcde} \}
~ \longrightarrow~ \{ \partial_{a} ,0,0 \}
\,,
\label{eq:solvesectionE6A}
\end{align}
thereby restricting the internal coordinate dependence to the first six coordinates, identically satisfies the section constraint (\ref{eq:section}). 
With this embedding, the Lagrangian
(\ref{eq:D5ExFT}) becomes equivalent to full $D=11$ supergravity. In turn, type IIB supergravity is recovered
upon choosing a second inequivalent solution of the section constraint based on the group decomposition
\begin{align}
\rm{E}_{6(6)} \supset~&    \mathbb{R}_{\rm IIB} \times {\rm SL}(5)\times{\rm SL}(2)
\,,\nonumber\\
{\bf 27} \longrightarrow ~&  (5,1)_{+4} \oplus (5',2)_{+1} \oplus (10,1)_{-2} \oplus (1,2)_{-5}
\,, \nonumber\\
\partial_M \longrightarrow ~& \{\partial_{i}, \partial^{i\alpha}, \partial^{ijk}, \partial_\alpha \} 
~ \longrightarrow~ \{\partial_{i}, 0,0,0 \} 
\,,
\label{eq:solvesectionE6B}
\end{align}
thereby restricting internal coordinate dependence to the first five coordinates.
With this embedding, the Lagrangian
(\ref{eq:D5ExFT}) becomes equivalent to full IIB supergravity.

The ExFT Lagrangian (\ref{eq:D5ExFT}) thus yields a common description of both, $D=11$ and IIB supergravity, depending on the choice of solution for the section constraint. Remarkably, by a slight but controlled violation of the section constraint, even the massive IIA theory can be included in the construction, see \cite{Hohm:2011cp,Ciceri:2016dmd,Cassani:2016ncu}.
The explicit dictionary between the ExFT fields (\ref{eq:ExFTfieldsE6}) and the fields of ten- and eleven-dimensional supergravity
has been worked out in \cite{Hohm:2013vpa,Baguet:2015xha}. For the 42 scalar fields parametrizing the matrix ${\cal M}_{MN}$, 
the embedding of the higher-dimensional fields is identical to the embedding in the torus reduction, 
discussed in section~\ref{sec:ungauged}. For example, the internal part $G_5$ of the IIB metric
can be straightforwardly extracted from the components of the inverse matrix ${\cal M}^{MN}$ according to
\begin{equation}
{\cal M}^{MN}\,\partial_M \otimes \partial_N =
({\rm det}\,{G_5})^{-1/3}\,G_5^{ij}\,\partial_i \otimes \partial_j
\,,
\label{eq:Mgint}
\end{equation} 
where the internal derivatives $\partial_i, \partial_j$ are 
embedded into the $\partial_M$ according to the solution (\ref{eq:solvesectionE6B})  of the section constraint.
The internal components of the IIB $p$-forms, are still identified as 
(\ref{eq:embedding_scalars}) within ${\cal M}_{MN}$. Indeed, in the ExFT formulation of the higher-dimensional supergravities,
the torus reduction is simply implemented by setting
\begin{equation}
\partial_M \longrightarrow0\,,
\end{equation}
upon which the Lagrangian (\ref{eq:D5ExFT}) straightforwardly reduces to (\ref{eq:D5}), 
i.e.\ to the maximal ungauged $D=5$ theory in the form in which the global ${\rm E}_{6(6)}$ is manifest.

Interestingly,  ExFT furthermore naturally accommodates many of the gauged supergravities discussed in section~\ref{sec:gauged}, and describes the embedding of the gauged Lagrangian (\ref{eq:D5gauged}) into higher dimensions. This is achieved via a generalized Scherk Schwarz reduction of the ExFT Lagrangian (\ref{eq:D5ExFT}), upon a proper extension of the standard Scherk-Schwarz reduction \cite{Scherk:1979zr} to the exceptional geometry. Explicitly, this reduction is described in terms of an E$_{6(6)}$-valued twist matrix $U$, and a weight factor $\rho$, both depending on the internal coordinates. The reduction ansatz for the external and the internal ExFT metric takes the factorized form
\begin{align}
\label{eq:SchSchgM}
g_{\mu\nu}(x,y) =\,& \rho^{-2}(y)\,{\rm g}_{\mu\nu}(x)\,,\nonumber\\
{\cal M}_{MN}(x,y) =\,& U_{M}{}^{\underline{K}}(y)\,U_{N}{}^{\underline{L}}(y)\,M_{\underline{KL}}(x)\,, 
\end{align}
where we now use underlined indices $\underline{M}$ to distinguish the $D=5$ fields from the ExFT fields.
Similarly, the reduction ansatz for the ExFT vector and tensor fields, takes the form
\begin{align}
{\cal A}_{\mu}{}^{M}(x,y) =\,&\rho^{-1}(y)\,(U^{-1})_{\underline{N}}{}^{M}(y)\, A_{\mu}{}^{\underline{N}}(x) \,,\nonumber\\
{\cal B}_{\mu\nu\,M}(x,y) =\,& 
\,\rho^{-2}(y)\, U_M{}^{\underline{N}}(y)\,B_{\mu\nu\,\underline{N}}(x)\,.
\label{eq:SchSchAB}
\end{align}
Consistency of the truncation Ansatz (\ref{eq:SchSchgM}), (\ref{eq:SchSchAB}) is encoded in a set of differential equations 
on the twist matrix 
\begin{equation}
\left[\Gamma_{\underline{MN}}{}^{\underline{K}}\right]_{{\bf 351}} = -\frac15\, X_{\underline{MN}}{}^{\underline{K}}
\,,\qquad \Gamma_{\underline{MN}}{}^{\underline{M}} =-4\,\rho^{-1}\partial_{\underline{N}} \rho 
\,,
\label{eq:SchSchConsistency}
\end{equation}
in terms of the algebra valued currents
\begin{equation}
\Gamma_{\underline{M}\underline{N}}{}^{\underline{K}}  \equiv 
\rho^{-1}\,(U^{-1})_{\underline{M}}{}^P\, (U^{-1})_{\underline{N}}{}^L\,\partial_P U_L{}^{\underline{K}} 
\,.
\label{Gamma}
\end{equation}
Here, 
$X_{\underline{MN}}{}^{\underline{K}}$ denotes the constant embedding tensor
characterizing the $D=5$ gauged theory, according to (\ref{eq:gg}). 
The projection $[\cdot]_{{\bf 351}} $ refers to the projection 
of the rank three tensor $\Gamma_{\underline{MN}}{}^{\underline{K}}$ 
onto the irreducible ${{\bf 351}}$ representation of E$_{6(6)}$ in which the embedding tensor transforms.
The quadratic constraint (\ref{eq:con2}) on this embedding tensor now follows as a consequence of the section constraint (\ref{eq:section}).
As a consequence of \eqref{eq:SchSchConsistency}, the ExFT field equations for (\ref{eq:SchSchgM}), (\ref{eq:SchSchAB})
factor into products of twist matrices and the field equations of the $D=5$ gauged theory (\ref{eq:D5gauged}), with embedding tensor $X_{\underline{MN}}{}^{\underline{K}}$.
In particular, the non-abelian field strengths (\ref{eq:FExFT}), (\ref{eq:HExFT}) reproduce the structures (\ref{eq:defF}), (\ref{eq:defH}) of the gauged supergravity.
It is useful to rephrase the consistency equations on the twist matrix \eqref{eq:SchSchConsistency} in terms of the global frame, ${\cal U}_{\underline{M}}{}{}^M\equiv \rho^{-1}(U^{-1})_{\underline{M}}{}{}^M$, as
\begin{equation}
{\cal L}_{{\cal U}_{{\underline{M}} }} {\cal U}_{\underline{N}}  =
X_{\underline{MN}}{}^{\underline{K}} \, {\cal U}_{\underline{K}} 
\,,
\label{eq:UUXU}
\end{equation}
where the action ${\cal L}_{{\cal U}_{{\underline{M}} }}$ of generalized diffeomorphisms is defined by \eqref{eq:genL} together with a canonical weight term. Mathematically, this equation states that the internal space is generalized Leibniz parallelizable \cite{Lee:2014mla}.

\begin{figure}[tb]
\center
\includegraphics[scale=.52]{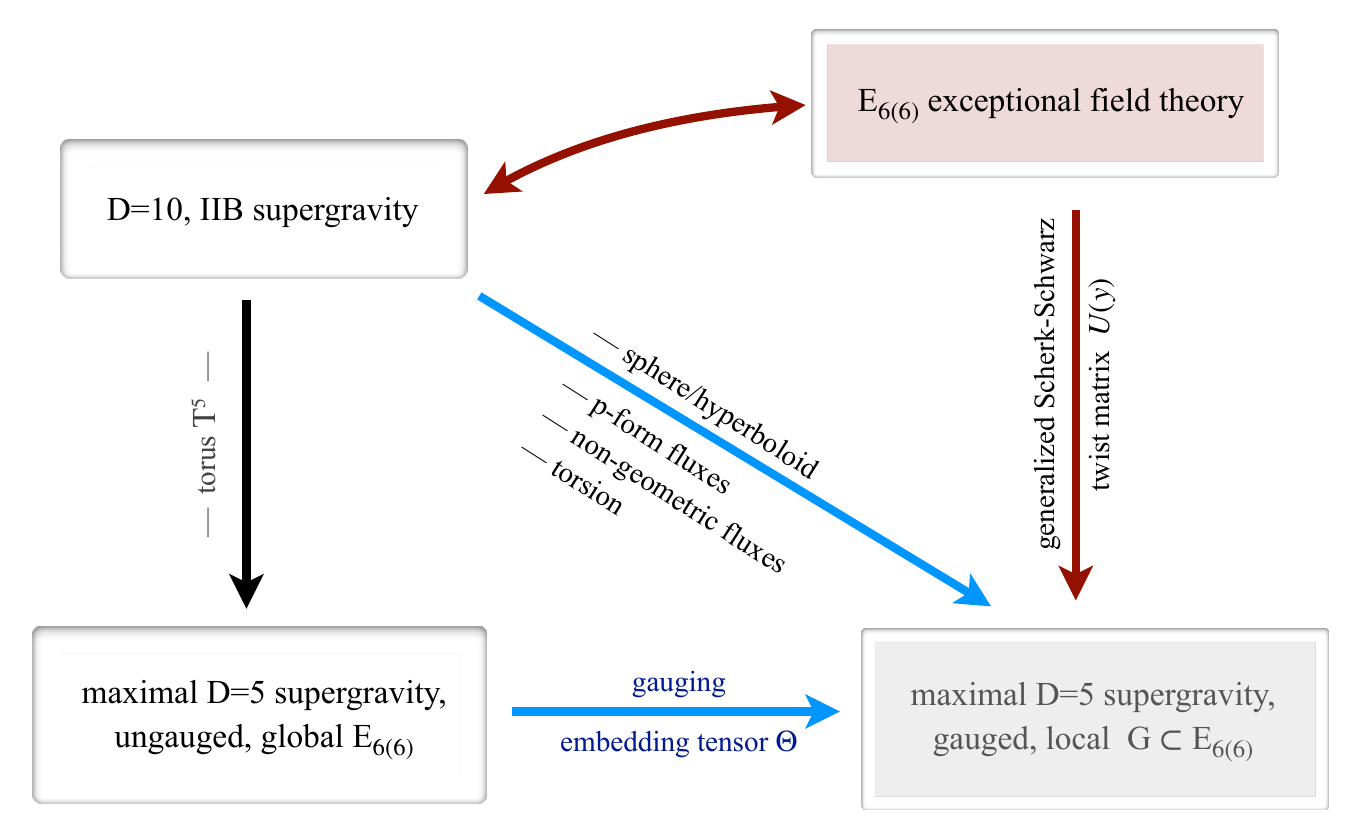}
\caption{
Exceptional field theory and consistent truncations}
\label{fig:exft}
\end{figure}

Every twist matrix solving \eqref{eq:SchSchConsistency} thus defines a consistent truncation of $D=11$ or IIB supergravity to a $D=5$ maximal supergravity in the strict sense that any solution to the $D=5$ field equations lifts to a solution of the higher-dimensional theory. Combining (\ref{eq:SchSchgM}) and (\ref{eq:SchSchAB}) with the dictionary between the ExFT fields (\ref{eq:ExFTfieldsE6}) and the fields of ten- and eleven-dimensional supergravity, yields the explicit embedding of the gauged maximal supergravity (\ref{eq:D5gauged}) into higher dimensions. 
We illustrate these connections in Figure~\ref{fig:exft}. Explicit twist matrices for sphere and hyperboloid reductions have been constructed in \cite{Lee:2014mla,Hohm:2014qga}. Other constructions include~\cite{Lee:2015xga,Inverso:2017lrz,DallAgata:2019klf,Galli:2022idq,Hassler:2022egz}. The ExFT framework thus offers a natural explanation for the consistent sphere truncations first established in \cite{deWit:1984nz} for AdS$_4 \times {S}^7$ and, among others, has allowed the construction of the analogous consistent truncation of IIB supergravity on AdS$_5 \times {S}^5$ \cite{Baguet:2015sma}. The formalism also allows to describe consistent truncations to theories with lower degree of supersymmetry, see \cite{Malek:2017njj,Cassani:2019vcl} and references therein.

Consistent truncations are an effective tool for the construction of solutions of the higher-dimensional theories by uplifting lower-dimensional solutions. As the simplest example, let us note that every stationary point of the scalar potential of a gauged supergravity defines a $D=5$ solution with constant scalars, vanishing vectors, and a maximally symmetric metric ${\rm AdS}_5$, ${\rm dS}_5$, or ${\rm Mink}_5$, depending on the value of the scalar potential. If the $D=5$ theory is obtained by consistent truncation, for instance, from IIB supergravity, this solution defines a ten-dimensional IIB solution of the type ${\rm AdS}_5\times {\Sigma}^5$ (in the case of a negative cosmological constant in $D=5$), where the metric on the internal space ${\Sigma}^5$ is defined by combining the embedding (\ref{eq:Mgint}) with the reduction ansatz (\ref{eq:SchSchgM}). An example of such a non-trivial IIB solution is the uplift \cite{Pilch:2000ej} of the ${\cal N}=2$ supersymmetric stationary point \cite{Khavaev:1998fb} in the ${\rm SO}(6)$-gauged supergravity.

Consistent truncations also provide a useful framework for performing reliable holographic computations in the lower-dimensional settings. In the final section, we will briefly review a particularly powerful application of  ExFT for the computation of Kaluza-Klein spectra around backgrounds living in these consistent truncations \cite{Malek:2019eaz}.

\section{Kaluza-Klein spectrometry}
\label{sec:KK}

As discussed so far, exceptional field theories provide the duality-covariant reformulation of higher-dimensional supergravity theories in a form specifically tailored to the description of the compactifications to lower dimensions. While the ExFT Lagrangian (\ref{eq:D5ExFT}) carries the full higher-dimensional theory, the reduction ansatz (\ref{eq:SchSchgM}), (\ref{eq:SchSchAB}) describes the embedding of the lower-dimensional theory (\ref{eq:D5gauged}). In terms of the Kaluza-Klein fluctuations around a given background, the lower-dimensional theory captures the non-linear dynamics of the lowest Kaluza-Klein multiplet, while the ExFT Lagrangian (\ref{eq:D5ExFT}) extends the dynamical equations to all states of the infinite Kaluza-Klein towers.

This description of the higher Kaluza-Klein fluctuations can be made explicit and gives rise to a very efficient description of those states, their masses and couplings within ExFT \cite{Malek:2019eaz,Malek:2020yue,Duboeuf:2023cth}. This is particularly useful in the context of holography, where these data translate into dimensions and correlation functions of operators in the dual field theory. Let us illustrate this for the ${\rm E}_{6(6)}$ ExFT discussed above. For a consistent truncation described by the reduction ansatz (\ref{eq:SchSchgM}), (\ref{eq:SchSchAB}), the higher Kaluza-Klein modes are captured by extending the ansatz to
\begin{align}
		g_{\mu\nu}(x,y) =\,& \rho^{-2}(y) \Big( {\rm g}_{\mu\nu}(x) + \sum_\Sigma {\cal Y}_\Sigma(y)\, h_{\mu\nu}{}^{\Sigma}(x) \Big) \,, \nonumber\\
		{\cal A}_\mu{}^M(x,y) =\,& \rho^{-1}(y)\, (U^{-1})_{\underline{M}}{}^M(y) \sum_\Sigma {\cal Y}_\Sigma(y)\, A_\mu{}^{{\underline{M}}\,\Sigma}(x) \,, \nonumber\\
		{\cal B}_{\mu\nu\,M}(x,y) =\,& \rho^{-2}(y)\, U_M{}^{\underline{M}}(y) \sum_\Sigma {\cal Y}_\Sigma(y)\, B_{\mu\nu\,{\underline{M}}}{}^{\Sigma}(x) \,,
		 \label{eq:KKAnsatz1}
\end{align}
for the spin-2, spin-1, and tensor fluctuations, as well as
\begin{equation}
	{\cal V}_M{}^{\underline{A}}(x,y) = U_M{}^{\underline{M}}(y)\,  \Big(\exp\Big(\tfrac12\, \mathbb{T}_{\underline{\alpha}} \sum_\Sigma {\cal Y}_\Sigma(y) \phi^{\underline{\alpha}\,\Sigma}(x) \Big)\Big){}_{\underline{M}}^{\;\;\;\;\underline{A}} \,,
	 \label{eq:KKAnsatz2}
\end{equation}
for the ${\rm E}_{6(6)}$-valued generalized vielbein encoding the generalized metric as ${\cal M}_{MN} ={\cal V}_{{\underline{M}}}{}^{\underline{A}}\, {\cal V}_{{\underline{N}}}{}^{\underline{B}}\, \delta_{\underline{A}\underline{B}}$.
Here, $\rho(y)$ and $U_M{}^{\underline{M}}(y)$ are the weight factor and the twist matrix, respectively, defining the truncation ansatz (\ref{eq:SchSchgM}), (\ref{eq:SchSchAB}). The $\mathbb{T}_{\underline{\alpha}}$ in (\ref{eq:KKAnsatz2}) refer to the non-compact generators of ${\rm E}_{6(6)}$, with the index $\underline{\alpha} = 1, \ldots, 42$, raised and lowered using the non-compact part of the ${\rm E}_{6(6)}$ Cartan-Killing metric. All explicit sums run over the scalar harmonics ${\cal Y}_\Sigma(y)$ on the internal space. For ${\rm SO}(6)$ supergravity, these would be the scalar harmonics on the round sphere $S^5$, organized into symmetric traceless vector representations ${\cal R}_{[n00]}$ of the isometry group ${\rm SO}(6)$.

The fact that only internal scalar harmonics appear in the expansions (\ref{eq:KKAnsatz1}), (\ref{eq:KKAnsatz2}), seems in striking contrast with the traditional Kaluza-Klein analysis, such as \cite{Kim:1985ez} for the AdS$_5\times S^5$ background. Typically, higher-dimensional fields in different representations of the Lorentz group on the internal space, require expansions into different types of internal harmonics \cite{Salam:1981xd}, corresponding to eigenfunctions of the different internal Laplacian operators. The simplification in the present structure is an illustration of the fact that the objects appearing under multiplication with the twist matrix in the fluctuation ansatz (\ref{eq:KKAnsatz1}), (\ref{eq:KKAnsatz2}) behave as singlets under the internal Lorentz group as a consequence of the consistency conditions (\ref{eq:SchSchConsistency}) on the twist matrix. Accordingly, their harmonic expansion is fully accounted for by the internal scalar harmonics. More explicitly, upon multiplying out the products among components of the twist matrix and the scalar harmonics in (\ref{eq:KKAnsatz1}), (\ref{eq:KKAnsatz2}), one recovers precisely the expected higher vector and tensor harmonics in the expansion of the IIB fields.

As a further consequence, one observes an intriguing structure of the full Kaluza-Klein spectrum: schematically, all fluctuations are naturally labeled in the form $\Phi^{{\cal A},\Sigma}$ with an index ${\cal A}$ referring to the field content of $D=5$ supergravity (or ExFT), and a second index $\Sigma$ labeling the scalar harmonics on the internal space, revealing a tensor product structure
\begin{equation}
\mbox{KK states} \longrightarrow \big(\mbox{$D=5$ supergravity}\big) \otimes \big(\mbox{scalar harmonics}\big)
\,.
\label{eq:KKtensor}
\end{equation}
It is important to stress that this factorized structure holds for the representations, but does not extend to include the Kaluza-Klein masses, and more generally applies to the un-Higgsed form of the spectrum in which all the Goldstone modes are included in the product on the r.h.s.\ of (\ref{eq:KKtensor}). Finally, let us note that for internal manifolds in the form of coset spaces, the product structure of (\ref{eq:KKtensor}) is precisely compatible with the group theoretical constraints defining the harmonic expansions according to \cite{Salam:1981xd}.

In order to obtain explicit mass formulas for the higher Kaluza-Klein states, it remains to plug the fluctuation ansatz (\ref{eq:KKAnsatz1}), (\ref{eq:KKAnsatz2}) into the linearized equations of motion of ${\rm E}_{6(6)}$ ExFT. The result can be expressed in terms of the embedding tensor $X_{\underline{MN}}{}^{\underline{K}}$ obtained from (\ref{eq:SchSchConsistency}), together with a constant matrix ${\cal T}_{\underline{M}}$ which captures the action of the generalized diffeomorphisms (\ref{eq:UUXU}) on the scalar harmonics as
\begin{equation}
	 {\cal L}_{{\cal U}_{\underline{M}}} {\cal Y}_{\Sigma} = - {\cal T}_{\underline{M}\,\Sigma\Omega}\, {\cal Y}_{\Omega} \,.
\end{equation}
In the rest of this section, we state the resulting Kaluza-Klein mass matrices, as obtained in \cite{Malek:2019eaz,Malek:2020yue,Duboeuf:2023cth}, for which we refer to details.
The mass matrix for the spin-2 fluctuations takes the simple Casimir-like form
\begin{equation} \label{eq:MM2}
	{\cal M}^{({\rm spin}\mbox{-}2)}_{\Sigma\Omega} = - {\cal T}_{\underline{M}\,\Sigma\Lambda}\, {\cal T}_{\underline{M}\,\Lambda\Omega} \,,
\end{equation}
as a product of the matrices ${\cal T}_{\underline{M}}$ acting on the scalar harmonics. This formula immediately follows from expanding the last terms in the ExFT potential (\ref{eq:potExFT}) with the fluctuation ansatz (\ref{eq:KKAnsatz1}). With the explicit embedding (\ref{eq:Mgint}) of the internal metric, it reproduces the general result of \cite{Bachas:2011xa} which shows that the mass spectrum of the spin-2 fluctuations is encoded in a massless scalar wave equation on the internal space.

 The vector mass matrix is given by
\begin{equation} \label{eq:MM1a}
	{\cal M}^{({\rm vector})}_{\underline{M}\,\Sigma,{\underline{N}}\,\Omega} = \frac{1}{12}\, \Pi_{\underline{M}\,\underline{\alpha}}{}_{\Sigma}{}_\Lambda\, \Pi^{\underline{\alpha}}{}_{{\underline{N}}}{}_{\Lambda}{}_{\Omega} \,,
\end{equation}
as a product of the matrices
\begin{equation}
	\Pi_{\underline{M}\,\underline{\alpha}}{}_\Sigma{}_\Omega = \Pi_{\underline{\alpha}\,\underline{M}\,\Omega}{}_{\Sigma} = - 2 \left( X_{\underline{MN}}{}^{\underline{K}}\,(\mathbb{T}_{\underline{\alpha}})_{\underline{K}}{}^{\underline{N}}\, \delta_{\Sigma\Omega} + 6\, (\mathbb{T}_{\underline{\alpha}})_{\underline{M}}{}^{{\underline{N}}}\, {\cal T}_{{\underline{N}\,}}{}_{\Sigma}{}_{\Omega} \right) \,,
\end{equation}
which describe the (linearized) coupling of vector to scalar fluctuations.
The masses of the tensor fluctuations are obtained from the antisymmetric matrix
\begin{equation}
	{\cal M}^{({\rm tensor})}_{\underline{M}\,\Sigma,{\underline{N}}\,\Omega} = \frac{1}{\sqrt{10}} \left( - Z^{\underline{M}{\underline{N}}} \delta_{\Sigma\Omega} + 10\, d^{\underline{M}{\underline{N}}{\underline{K}}}\, {\cal T}_{{\underline{K}}\,\Sigma\Omega} \right) \,,
	\label{MM1b}
\end{equation}
where we recall that the tensor fluctuations, just as the tensor fields in the $D=5$ theory, satisfy a first-order equation (\ref{eq:D5HB}) of topologically massive fields, obtained from expanding (\ref{eq:ExFTFH}). 
Finally, the scalar mass matrix is obtained from expanding the ExFT potential (\ref{eq:potExFT}), and can be given in the rather compact form
\begin{align} \label{eq:MM0}
		M^{({\rm scalar})}_{\underline{\alpha}\,\Sigma,\underline{\kappa}\,\Omega}  
		=\;& \mathbb{M}_{\underline{\alpha}\underline{\kappa}}\, \delta_{\Sigma\Omega} 
		+ 2\,\mathbb{N}_{\underline{M}\,}{}_{\underline{\alpha}\underline{\kappa}} {\cal T}_{\underline{M}\,}{}_{\Sigma\Omega} + \delta_{\underline{\alpha}\underline{\kappa}} {\cal M}^{({\rm spin}\mbox{-}2)}_{\Sigma\Omega} 
		 \nonumber\\
		&
		- \tfrac1{12} \Pi_{\underline{\alpha}\,}{}_{\underline{M}}{}_{\Lambda}{}_{\Sigma} \, \Pi_{\underline{M}\,}{}_{\underline{\kappa}}{}_{\Lambda}{}_{\Omega}\,,
\end{align}
where the last term only affects the (non-physical) Goldstone modes, and the other terms are given by
\begin{align} \label{eq:MassMatrixScalar12}
		\mathbb{M}_{\underline{\alpha}\underline{\kappa}} =\,& X_{\underline{M}\underline{K}}{}^{\underline{L}} X_{\underline{N}\underline{L}}{}^{\underline{K}} \, (\mathbb{T}_{\underline{\alpha}}\mathbb{T}_{\underline{\kappa}})_{\underline{M}}{}^{\underline{N}} +  \tfrac12 X_{\underline{M}\underline{K}}{}^{\underline{L}} X_{\underline{N}\underline{K}}{}^{\underline{L}}\,  (\mathbb{T}_{\underline{\alpha}}\mathbb{T}_{\underline{\kappa}})_{\underline{M}}{}^{\underline{N}} 
		\nonumber\\
		& \quad + X_{\underline{M}\underline{K}}{}^{\underline{P}} X_{\underline{N}\underline{L}}{}^{\underline{P}}\, 
		(\mathbb{T}_{\underline{\alpha}})_{\underline{M}}{}^{\underline{N}}\, (\mathbb{T}_{\underline{\kappa}})_{\underline{K}}{}^{\underline{L}} 
		+ \tfrac13 X_{\underline{M}\,\underline{\alpha}} X_{\underline{M}\,{\underline{\kappa}}} \,, \nonumber\\[1ex]
		\mathbb{N}_{\underline{M}\,\underline{\alpha}\underline{\kappa}} =\,& 
		\mathbb{N}_{\underline{M}}{}^{\hat{\alpha}} \,  (\mathbb{T}_{\hat{\alpha}})_{\underline{\alpha}\underline{\kappa}}
				= 
				\left( X_{{\underline{M}}}{}^{\hat{\alpha}} + 3 X_{\underline{K}\underline{L}}{}^{{\underline{M}}}\, (\mathbb{T}^{\hat{\alpha}}){}_{\underline{K}}{}^{\underline{L}} \right)
				(\mathbb{T}_{\hat{\alpha}})_{\underline{\alpha}\underline{\kappa}}		\,.
\end{align}
Here, $\mathbb{M}_{\underline{\alpha}\underline{\kappa}} $ is the mass matrix of $D=5$ supergravity, and the $\mathbb{N}_{\underline{M}}$ denote a set of compact generators, with indices $\hat\alpha=1, \dots, 36$, labeling the compact generators $\mathbb{T}_{\hat{\alpha}}$ within ${\rm E}_{6(6)}$.

The mass matrices (\ref{eq:MM2})--(\ref{eq:MM0}) express the full Kaluza-Klein spectrum in terms of the embedding tensor of the $D=5$ supergravity, and the matrix ${\cal T}_{\underline{M}}$ acting on the scalar harmonics. For the AdS$_5\times S^5$ background, they reproduce in compact form the result originally obtained in \cite{Kim:1985ez} by linearizing the IIB field equations around this background. The universal form of the mass matrices is a consequence of the covariance of ExFT under the exceptional symmetry groups. As a result, 
these very same matrices are likewise applicable for computing the spectrum around any other vacuum of the $D=5$ theory, simply upon changing the associated twist matrix according to
\begin{equation}
	U_M{}^{{\underline{M}}} \,\longrightarrow\, U_{M}{}^{\underline{A}} = U_M{}^{{\underline{M}}}\, \mathring{{\cal V}}_{{\underline{M}}}{}^{\underline{A}} \,,
\end{equation}
where the ${\rm E}_{6(6)}$ matrix $\mathring{{\cal V}}$ describes the scalar fields at the vacuum within the $D=5$ theory. This structure has been exploited, for example, in \cite{Bobev:2020lsk} in order to determine in closed form the full Kaluza-Klein spectrum around the ${\rm U}(2)$ deformed $S^5$ background of \cite{Pilch:2000ej}, corresponding to the ${\cal N}=2$ stationary point of ${\rm SO}(6)$-gauged supergravity. In particular, this spectrum features infinite towers of both long and protected multiplets, of which the latter precisely match  the superconformal index computed on the dual SCFT side. In the same fashion, the above mass matrices can be evaluated for the IIB vacua associated with all other stationary points of the ${\rm SO}(6)$ scalar potential, found in \cite{Bobev:2020ttg}, which break all supersymmetries and preserve very little to none of the bosonic ${\rm SO}(6)$ symmetries. Unfortunately, all these vacua are unstable already within $D=5$ supergravity.
Similar applications of ExFT for the computation of Kaluza-Klein spectra have been worked out in many other examples and dimensions, including perturbatively stable non-supersymmetric AdS vacua, with AdS$_4$ vacua in massive IIA supergravity~\cite{Guarino:2020flh}, and AdS$_3$ vacua in IIB supergravity~\cite{Eloy:2023zzh}. An extension of the formulas for the mass matrices for the fermionic fluctuations has been spelled out in \cite{Cesaro:2020soq}. Furthermore, similar universal expressions may be derived from ExFT for cubic and higher couplings among the Kaluza-Klein states \cite{Duboeuf:2023cth}.

\section{Conclusions}
\label{sec:conclusions}

We have briefly reviewed exceptional field theories as the duality-covariant reformulation of maximal supergravity theories in ten and eleven dimensions, that make the underlying exceptional symmetries explicit.
Beyond their foundational role in unifying the various maximal supergravities, we have illustrated how they also provide access to very efficient techniques for tackling concrete computational problems in supergravity.
To conclude, it is worth noting that although our discussion here has focused on the bosonic sector of maximal supergravities, exceptional field theories can be uniquely extended to the fermionic sector, such that the full higher-dimensional supersymmetry is realized on the ExFT fields \cite{Coimbra:2012af,Godazgar:2014nqa,Musaev:2014lna,Butter:2018bkl,Bossard:2019ksx}. However, as noted above, supersymmetry is no longer necessary to constrain and determine the couplings of the bosonic sector, which have been fixed uniquely by imposing purely bosonic symmetries, specifically the generalized internal and external diffeomorphisms. 
In this sense, the exceptional symmetries have replaced supersymmetry as a defining principle of maximal supergravities.
This may spark the hope of further intriguing consequences of these structures, in particular extending to the quantum level of maximal supergravities.

\bigskip

\noindent
{\em Acknowledgements:}
I wish to thank all my collaborators and in particular Arnaud Baguet, Guillaume Bossard, Franz Ciceri, Bastien Duboeuf, Camille Eloy,
Bernard de Wit, Olaf Hohm, Gianluca Inverso, Axel Kleinschmidt, Emanuel Malek, Hermann Nicolai, 
Ergin Sezgin, and Mario Trigiante, for all the numerous exciting 
discussions and projects --- towards, within, and beyond exceptional field theory.


\providecommand{\href}[2]{#2}\begingroup\raggedright\endgroup

\end{document}